\documentclass[aps,pre,twocolumn,groupedaddress,showpacs,floatfix,superscriptaddress,longbibliography]{revtex4-2}
\usepackage{amsmath}
\usepackage{amssymb}
\usepackage{physics}
\usepackage{graphicx}
\usepackage{tabularx}
\usepackage[dvipsnames]{xcolor}
\usepackage[normalem]{ulem}
\usepackage{soul} 
\usepackage{ragged2e}
\usepackage{hyperref}
\usepackage{booktabs}
\usepackage{orcidlink}

\usepackage{tikz}
\usetikzlibrary{calc}

\usepackage[caption=false]{subfig}
\MakeRobust{\subref}

\emergencystretch=\maxdimen
\hypersetup{
 colorlinks=true,
 linkcolor=blue,
 anchorcolor = blue,
 citecolor = blue,
 filecolor = blue,
 urlcolor = blue
}

\newcommand{\oist}{Okinawa Institute of Science and Technology Graduate University, Onna-son, Okinawa 904-0495, Japan}

\newcommand{\unesp}{São Paulo State University (UNESP), Institute of Chemistry, 14800-090, Araraquara, São Paulo, Brazil}

\newcommand{\uspsc}{Instituto de Física de São Carlos, Universidade de São Paulo, CP 369, 13560-970 São Carlos, São Paulo, Brazil}

\newcommand{\etal}[1]{\textit{et al} }

\begin{document}
\title{Textures as a phase-transition probe for quantum spin chains}

\author{Heitor P. Casagrande\,
\orcidlink{0000-0003-0247-339X}}
\email{heitor-peres@oist.jp}
\affiliation{\oist}

\author{Isaac M. Carvalho,
\orcidlink{0000-0002-0117-3283}}
\affiliation{\unesp}

\author{William J. Munro,
\orcidlink{0000-0003-1835-2250}}
\affiliation{\oist}

\author{Krissia Zawadzki\,
\orcidlink{000-0002-6133-0850}}
\affiliation{\uspsc}




\begin{abstract}

The idea of quantum texture has been recently proposed \cite{PhysRevLett.133.260801} and used as a tool for quantifying coherences and for quantum gate identification. In this work we offer a study on its usage to quantum phase transitions, demonstrating the rugosity metric as a simple tool for effective phase-transition probing. We establish the link between rugosity in the computational basis and the hierarchy of spin correlators, and analyze rugosities defined in the global ground-state and in ground-states belonging to different magnetization sectors (to which we refer to as global vs  symmetry-resolved rugosities) to study the phase diagram of the Heisenberg XXZ model. We find distinct rugosity signatures at both transition points. In particular, a sharp feature appears at $\Delta=1$ already for small systems, revealing a pronounced sensitivity of the correlation hierarchy encoded by the texture to this point. Since the BKT transition coincides with the isotropic $SU(2)$ point of the XXZ model, this behavior may reflect a particular sensitivity of rugosity to the structure of the spin-correlation hierarchy at isotropy.
\end{abstract}

\maketitle

\setcounter{figure}{0}


\section{Introduction}

Quantum phase transitions (QPTs) in one-dimensional (1D) systems provide one of the most striking manifestations of the enhanced role of quantum fluctuations in low dimensions~\cite{giamarchi2003quantum,gogolin1999,Sachdev_1999}. In such systems, quantum fluctuations can become sufficiently strong to suppress long-range order, preventing the spontaneous breaking of continuous symmetries under broad conditions, as established by the Mermin--Wagner--Hohenberg theorem~\cite{PhysRev.158.383}. Consequently, many QPTs involving continuous symmetries, such as $SU(2)$ spin-rotation  symmetry or $U(1)$ charge conservation symmetry, evade conventional Landau paradigm, based on local order parameters that characterize the spontaneous symmetry breaking. Such QPTs belong to a broad class of unconventional phenomena~\cite{gogolin1999,Chen2025,YU20261} and have long motivated the search for alternative descriptors of phase transitions that provide physical insight while remaining experimentally and numerically accessible.

However, in the absence of a suitable local order parameter, identifying reliable signatures of unconventional QPTs often requires observables capable of capturing the underlying physics. Alternatively, one may turn to full-state tomography~\cite{Hauke2016}, a significant and generally impractical undertaking in many-body systems because of its exponential resource requirements~\cite{Cramer2010}. Several complementary approaches have been employed to characterize QPTs, including entanglement-based diagnostics, which reveal the entanglement structure and universal features of critical systems~\cite{Osterloh2002,PhysRevA.66.032110,calabrese2007entanglement,Calabrese_2004}; machine-learning methods for identifying and classifying phase transitions directly from data~\cite{Franco2026,peleteiro2026learningspectraldensityfunctions,rattighieri2026measurementguidedstaterefinementshallow}; Bell-inequality-based methods for probing nonclassical many-body correlations~\cite{Justino2012}; and, more recently, measures of quantum nonstabilizerness~\cite{nehra2025}.

A paradigmatic setting for exploring unconventional quantum criticality is provided by the Berezinskii--Kosterlitz--Thouless (BKT) transition~\cite{Berezinskii1971,Kosterlitz1973}. Unlike conventional Landau transitions, BKT criticality is governed by a topological mechanism rather than by spontaneous symmetry breaking and the emergence of a local order parameter. A distinctive hallmark of this transition is the exponential, rather than power-law, closing of the excitation gap as the critical point is approached. Consequently, the correlation length grows extremely rapidly near criticality, requiring increasingly larger systems to reach the asymptotic critical regime. As a result, critical signatures often remain obscured in experimentally and numerically accessible system sizes, making the BKT transition particularly challenging to characterize and reliable probes of criticality especially valuable.

Current literature into the BKT transition discrimination relies on gap scaling~\cite{Dalmonte2015}, requiring both large system sizes and high sweeping resolution in order to be ascertained. 
Among entanglement witnesses, coherence is frequently used as well~\cite{RevModPhys.89.041003}: single spin coherence measures have been shown to detect quantum phase transitions where pair-wise entanglement measures fail~\cite{PhysRevB.90.104431, camak2015}, and crucially, coherence is naturally linked to the full structure of correlations in the state through its connection to off-diagonal elements in the density matrix, allowing for a more global witness than bipartite concurrence \cite{PhysRevLett.113.140401,RevModPhys.89.041003}. A simple measurable quantity, sensitive to criticality, would be extremely valuable, particularly if it allows for insight into the system dynamics with limited number of measurements \cite{Huang2020}.

Recently, the concept of Quantum Textures has been put forward~\cite{PhysRevLett.133.260801, PRAroleOfQuantumStateTexture}, and has created increasing interest in the quantum information community. In the seminal work, Par\'isio proposes a texture quantifier called \textit{rugosity}, that captures information about quantum coherence without requiring complete state reconstruction. Notably, the author was able to develop a quantum-gate identification procedure by leveraging this formalism.

Here, we present a study of quantum textures formalism to condensed matter physics, extending the textures framework in a completely different direction: we demonstrate a simple, computationally efficient probe of quantum-phase transition, not only sensitive to the two phases present in the XXZ model, but that also provides different scalings and signatures, characterizing the phase transitions across both points.

Rugosity~\cite{PhysRevLett.133.260801}, as we will show, fills this role by encoding a hierarchy of multi-point spin correlation functions in the texture, positioning it as a more powerful witness than pairwise entanglement measures or single-spin correlators.
Additionally, we show rugosity upper bounds global entanglement. This is established by writing rugosity in terms of two site-correlators for parity-symmetric states, showing that, for block diagonalizable systems, one is able to probe the two-site correlators using the global texture as a proxy.

\section{Quantum Textures}\label{textureBackground}
We consider a quantum state $\hat{\rho}$ represented in a fixed basis $\{ \ket{i} \}$ of dimension $D$. This state can be represented as a matrix of dimension $D \times D$ and plotted as a 3D histogram where each entry takes the real or the imaginary part of the matrix elements $\rho_{ij} = \braket{i| \hat{\rho}|j}$. The landscape of this histogram usually has some \textit{texture} whenever the heights of the bars differ. A uniform or \textit{textureless} landscape on the other side, coincides with a table which contains all heights equal to $1/D$. There is a single state that translates into an uniform histogram,
\begin{equation}
    \ket{f_1} = \frac{1}{\sqrt{D}} \sum_{i=1}^D \ket{i}.
    \label{eq:f1}
\end{equation}
The state $\ket{f_1}$ is \textit{textureless} and corresponds to the zero-resource set of the theory. It has a uniform distribution of elements across all matrix elements, making it quite a rich state. For a quantum state $|\hat{\rho}\rangle$, there are two quantities of interest: texture, and rugosity. Texture is given by
\begin{equation}
    \mathcal{T} = D \langle f_1| \hat{\rho} | f_1 \rangle
    \label{eq:Texture}
\end{equation}
While rugosity is defined as
\begin{equation}
    \mathcal{R} = -\ln \langle f_1| \hat{\rho} | f_1 \rangle = -\ln \left( \text{Tr}\left( | f_1 \rangle  \langle f_1| \hat{\rho} \right) \right)
    \label{eq:Rugosity00}
\end{equation}
In a fixed measurement basis (here we use the computational basis), $\mathcal{R}$ quantifies the texture of the state. Rugosity thus captures inhomogeneities that signal underlying system structure or change, including phase transitions, as we will show. We note that, since textures are defined in term of a fixed basis $\{ \ket{i} \}$, rugosity is basis-dependent.

\subsection{Texture correlation hierarchy in 2-level systems}

Initially, we consider a general system composed of $N$ 2-level sub-systems $\{ |g\rangle, |e\rangle\}$, yielding a Hilbert-space dimension of $d=2^N$. The reference textureless state defined from Eq.~\ref{eq:f1} can be re-factorized as
\begin{equation}
    \ket{f_1}
    =
    \bigotimes_{j=1}^{N}\ket{+}_j,
\end{equation}
where $\ket{+} = \frac{1}{\sqrt{2}}\left(\ket{g}+\ket{e}\right)$. For an arbitrary density matrix $\hat\rho$, the texture $\mathcal{T}_N = \bra{f_1}\rho\ket{f_1}$ can be written as the following
\begin{equation}
    \mathcal{T}_N
    =
    \frac{1}{2^N}
    \left\langle
        \prod_{j=1}^{N}(1+\sigma_j^x)
    \right\rangle,
\end{equation}
where $\sigma_j^x$ is the Pauli-x operator, corresponding to the NOT gate in quantum computing, and $\langle O\rangle = \Tr(\rho O)$.

Expanding the product, we find
\begin{align}
    \mathcal{T}_N
    =
    \frac{1}{2^N}
    \Bigg[
    1
    &+
    \sum_i \langle \sigma_i^x\rangle
    +
    \sum_{i<j}\langle \sigma_i^x\sigma_j^x\rangle
    +
    \sum_{i<j<k}\langle \sigma_i^x\sigma_j^x\sigma_k^x\rangle
    \nonumber\\
    &+
    \sum_{i<j<k<l}\langle \sigma_i^x\sigma_j^x\sigma_k^x\sigma_l^x\rangle
    +\cdots+
    \left\langle\prod_{j=1}^{N}\sigma_j^x\right\rangle
    \Bigg].
\end{align}
Although this is general, depending on the system's underlying symmetries, even or odd correlators will vanish. This will be further studied in Sec.~\ref{sec:globalAndSymmetryResolved}, where we study the rugosities of systems with fixed magnetization sectors. Now we turn to our system of interest.
\section{Phase transitions across the spin 1/2 XXZ model}

Initially, we study the Heisenberg spin-1/2 XXZ Hamiltonian on a chain of length $N$, given by
\begin{multline}
    H = \sum_{l=1}^{N-1} \left( J_x \sigma^{x}_{l} \sigma^{x}_{l+1} + J_y \sigma^{y}_{l} \sigma^{y}_{l+1} + \Delta \sigma^{z}_{l} \sigma^{z}_{l+1}\right) \\
    + \sum_{l=1}^{N} \left( h_x \sigma^{x}_{l} + h_y \sigma^{y}_{l} + h_z \sigma^{z}_{l}\right),
    \label{eq:Ham}
\end{multline}
with $J_x = J_y = 1$ and the variable $\Delta$ along the z-axis, and tunneling fields $h_x,~h_y,~h_z$. This particular model presents two zero-temperature phase transitions: a first-order transition at $\Delta = -1$ between the ferromagnetic ($\Delta < -1$) and gapless XY ($-1 < \Delta < +1$) regimes, and Berezinskii--Kosterlitz--Thouless (BKT) transition (sometimes referred to as infinite order) at $\Delta = +1$, between the gapless XY and Néel antiferromagnet ($\Delta > +1$) \cite{Kosterlitz1973, Haldane1980, Wang2010}. Conventional probes such as staggered magnetization are able to capture the latter, but not the former \cite{Mikeska2004}. A more reliable measure shown to signal both are Bell measurements~\cite{Justino2012}, but at the cost of requiring many realizations for the Bell Inequality violation calculation, which are generally cumbersome. Our results on the other hand establish the link between the nature of such metrics, stating rugosity as a proxy measure for entanglement that is also sensitive to site-site correlators, while depending only on simple measurements.

To showcase this, we employ Density Matrix Renormalization Group (DMRG) \cite{PhysRevLett.69.2863, fishman2022itensor} simulations as well as Exact Diagonalization (ED) analysis to report findings for multiple systems with different conditions in order to show the full scope of the power of textures when analyzing quantum phase-transitions. We focus on the XXZ Hamiltonian. We set therefore $J_x = J_y = 1$, and keep $\Delta$ variable. To break any degeneracies within the system we add a small staggered magnetic field $h_z = \pm 10^{-5}$ contingent on the site index (odd sites are under $-h_z$). All simulations were run until convergence under error $\epsilon = 10^{-6}$.

\section{Results}\label{Results}

\subsection{Global and symmetry-resolved rugosity} \label{sec:globalAndSymmetryResolved}
It is insightful to understand how the rugosity measure interacts with the system by defining three distinct measures: global, local and restricted to the a given symmetry sector defined in terms of quantum number $Q$, to which we refer as symmetry-resolved rugosity $\mathcal{R}_Q$. For the XXZ chain, the symmetry-resolved rugosity is built by restricting the system to a certain magnetization sector. We note this requires block-diagonalizing the system, which is not always feasible, but is nevertheless insightful for this discussion.
\begin{figure}
    \centering
    \includegraphics[width=1.0 \linewidth]{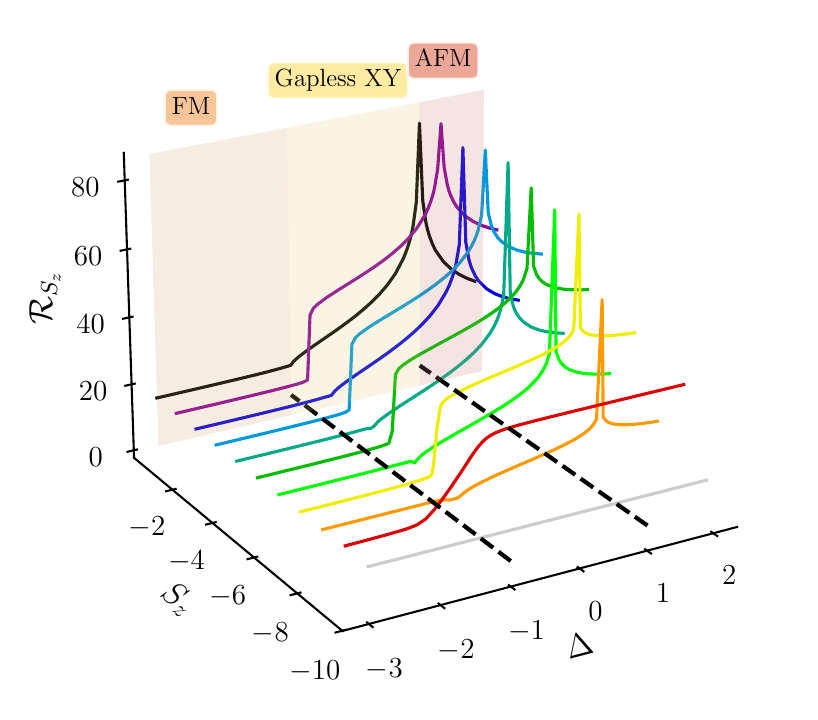}
    \caption{Symmetry-resolved rugosity $\mathcal{R}_{S_z}$ versus $\Delta$ for different magnetization sectors. The parity dependence of the symmetry-resolved rugosity sectors can be clearly seen at the point $\Delta = -1$, but the peak at $\Delta = +1$ is present at every sector but the fully polarized ones.}
    \label{fig:localRugosity}
\end{figure}
We demonstrate the symmetry-resolved character of rugosity in Fig~\ref{fig:localRugosity} by showing the rugosity behavior for different magnetization sectors. We see a few important features of rugosity, including a clear parity effect taking shape: although a sharp signature occurs at $\Delta=1$, the behavior at the first-order transition depends on whether the magnetization sector is odd or even.
In contrast for a global texture, one must simply not constrain the system. The global texture collapses the symmetry-resolved rugosity curves into a single one. Since the rugosity as defined in \cite{PhysRevLett.133.260801} is additive, this is supportive of the behavior. That is not to say that the global rugosity does not offer insight into the dynamics, as we will soon show.

\subsection{Global Rugosity}

\begin{figure}
    \centering
    \includegraphics[width=1.0 \linewidth, trim={0.5cm 2.1cm 0 1cm}, clip]{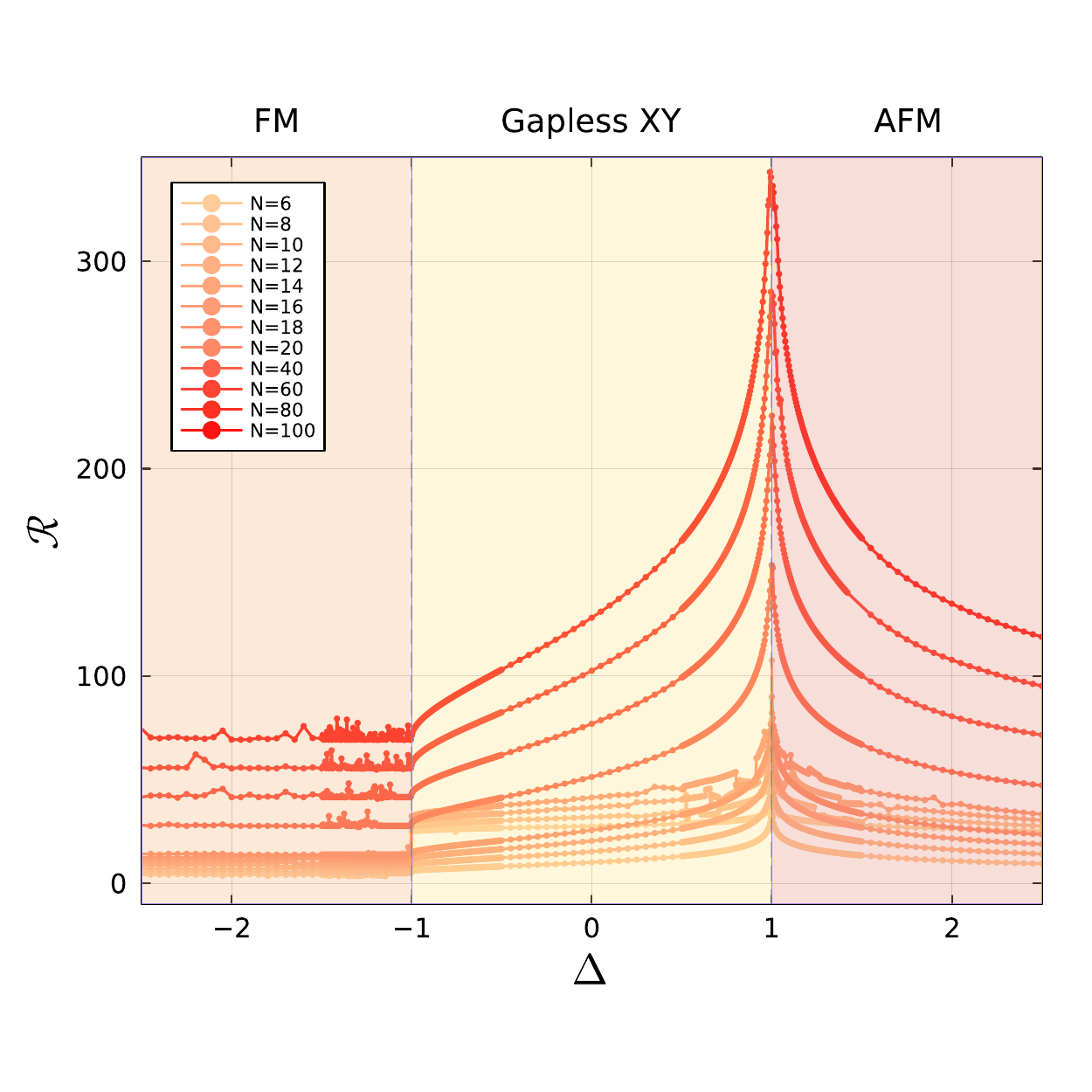}
    \caption{Global rugosity $\mathcal{R}$ versus anisotropy $\Delta$ for different system sizes. Lighter tones correspond to smaller sizes, and darker tones to larger sizes. We see a consistent trend among all system-sizes: a slow increase starting at $\Delta=-1$ and a pronounced peak at $\Delta=1$.}
    \label{fig:rugosityVsJz}
\end{figure}

The global rugosity can be seen in Fig.~\ref{fig:rugosityVsJz}. We see that the rugosity parameter is sensitive to both phase transition points, showing a sharp jump at $\Delta = -1$ and a sharp peak at $\Delta = +1$. Remarkably, the position of the latter remains pinned at $\Delta=1$ already for the smallest chains considered and shows no discernible finite-size shift as the system size increases. We note that the BKT transition coincides with the isotropic point of the XXZ model, where, in the absence of the small symmetry-breaking field, the anisotropy between the longitudinal and transverse spin couplings vanishes and the Hamiltonian acquires full $SU(2)$ symmetry. Since the texture encodes a hierarchy of transverse spin correlations, this point is particularly relevant: at $\Delta=1$, the distinction between longitudinal and transverse spin directions disappears. The sharp rugosity feature may therefore reflect a particular sensitivity of this hierarchy of correlations to the isotropic point, possibly explaining the absence of the finite-size shift typically expected near a BKT transition.

Beyond the position of these features, the rugosity also exhibits distinct behaviors across the different regimes, being mostly constant at $\Delta \leq -1$, and presenting a steady increase at $-1 \leq \Delta < +1$. A fast decrease then follows beyond $\Delta=+1$. To further understand the size dependence of these features, we analyze the scaling behavior at both transition points, as shown in Fig.~\ref{fig:rugosity_scaling_fits}. We see that the BKT transition presents a much sharper increase in rugosity as the thermodynamic limit $1/N \rightarrow 0$ is approached. The corresponding scaling fits further quantify this trend.

\subsection{Rugosity size-scaling analysis}

\begin{figure}
    \centering

\begin{tikzpicture}
    \node[] (fig) at (0,0){\includegraphics[width=0.8\linewidth, trim={0.8cm 0.cm 0 0},clip]{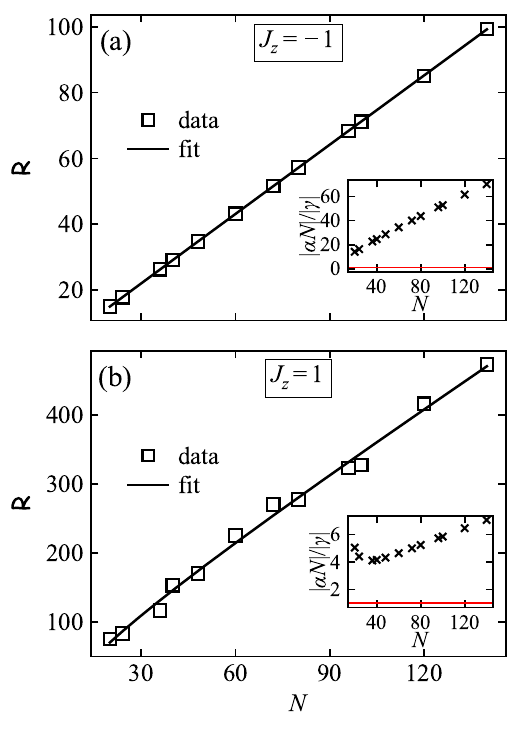}};
    \node[anchor=north, font=\large, rotate=90] (fig-t1) at ($(fig.west)+(-0.35cm,-2cm)$){$\mathcal{R}$};
    \node[anchor=north, font=\large, rotate=90] (fig-t2) at ($(fig.west)+(-0.25cm,3cm)$){$\mathcal{R}$};  

    \node[anchor=north, font=\large, rectangle, fill=white, inner sep=2mm] (fig-t3) at ($(fig.north)+(0.125 cm,-0.4cm)$){$\Delta = -1$}; 

    \node[anchor=north, font=\large, rectangle, fill=white, inner sep=2mm] (fig-t4) at ($(fig.north)+(0.125 cm,-5.3cm)$){$\Delta = 1$};

\end{tikzpicture}
    
    \caption{Finite-size scaling of the rugosity at the phase-transition points $\Delta=\{-1,1\}$. Symbols denote numerical data, while solid lines correspond to fits using $\mathcal{R}(N)=\alpha N+\beta\log N+c$. Insets show the ratio $|\alpha N|/|\gamma|$, with $\gamma=\beta\log N+c$; the red horizontal line marks $|\alpha N|/|\gamma|=1$, where extensive and sub-leading contributions have equal magnitude.}
    \label{fig:rugosity_scaling_fits}
\end{figure}
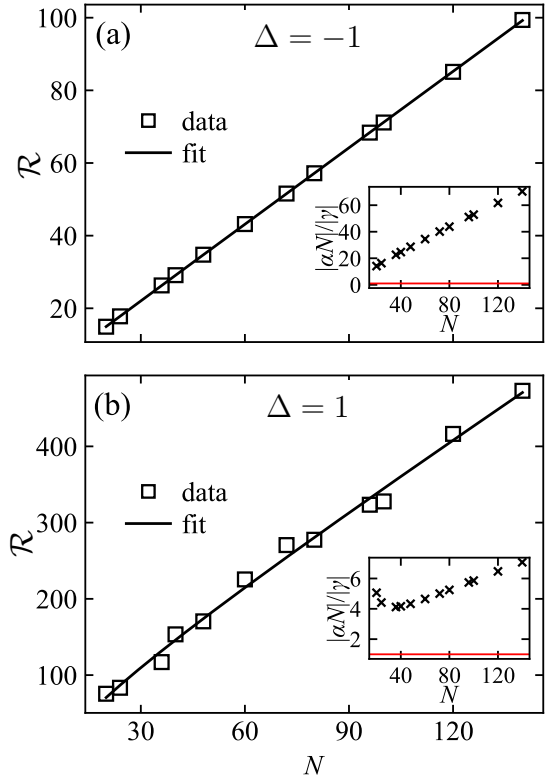

To gain further insight into the finite-size structure of the rugosity, we analyze its scaling with system size. As a global quantity associated with the spatial structure of the quantum state, the rugosity is expected to exhibit a leading contribution that increases with system size, supplemented by finite-size corrections. We therefore consider the scaling ansatz
\begin{equation}
\mathcal{R}(N)=\alpha N+\gamma,
\label{eq_scaling}
\end{equation}
where $\alpha$ denotes the leading extensive contribution and
$\gamma=\beta\ln N+c$ collects the subleading corrections.

As shown in Fig.~\ref{fig:rugosity_scaling_fits}, the scaling ansatz provides a good description of the numerical data at both transition points, despite the enhanced finite-size effects expected near the BKT transition at $\Delta=1$. To quantify the relative importance of the leading and subleading contributions, we consider the ratio $|\alpha N|/|\gamma|$. By construction, $|\alpha N|/|\gamma|=1$ corresponds to the point at which the extensive and subleading sectors contribute equally to the scaling behavior, while larger values indicate an increasing dominance of the extensive term. The insets show that this ratio remains above unity and increases with system size in both cases. In particular, values of $|\alpha N|/|\gamma|\gtrsim 4$ near the BKT transition and $|\alpha N|/|\gamma|\gtrsim 15$ at $\Delta=-1$ imply that the extensive contribution accounts for more than $80\%$ and $94\%$ of the rugosity, respectively.  By contrast, for entanglement entropies the hallmark of criticality is provided by the logarithmic scaling governed by the conformal-field-theory central charge~\cite{Calabrese_2004,Amico2008}. For texture, however, the present results show that its finite-size scaling is predominantly controlled by the extensive sector. Importantly, the relevant diagnostic is the behavior of the extensive coefficient $\alpha$ across the transition, rather than its dominance alone. The subleading sector may still contain additional information about the finite-size structure of the texture; however, within the system sizes considered here, its behavior does not allow us to establish a universal logarithmic contribution associated with BKT criticality.

\begin{figure}
\begin{tikzpicture}
    \node[] (fig) at (0,0){\includegraphics[width=0.8\linewidth, trim={0cm 0.8cm 0 0},clip]{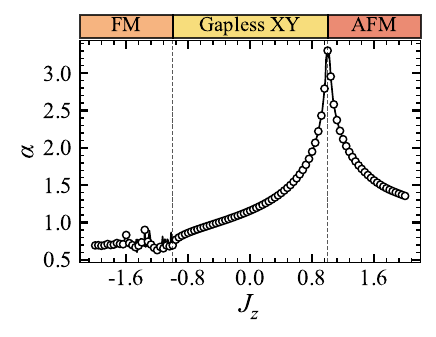}};
    \node[anchor=north, font=\large] (fig-t) at ($(fig.south)+(0.5cm,0)$){$\Delta$};
\end{tikzpicture}

    \caption{Extensive coefficient $\alpha$ extracted from fits of the form $\mathcal{R}(N)=\alpha N+\beta\log N+c$ as a function of $\Delta$. Vertical dashed lines indicate the transition points at $\Delta=\pm1$.}
    \label{fig:alpha_vs_Jz}
\end{figure}

We therefore turn to the behavior of the coefficient $\alpha$ in Eq.~\eqref{eq_scaling} across the phase diagram. Figure~\ref{fig:alpha_vs_Jz} reveals a pronounced global maximum of $\alpha$ around $\Delta=1$, whereas no comparable feature is observed near $\Delta=-1$. Since the subleading corrections remain comparatively small throughout the system-size range considered, the rugosity is well approximated by $\mathcal R\sim \alpha N$. The profile of $\alpha$ thus mirrors the behavior of rugosity, reproducing its most prominent feature around $\Delta=1$, where the BKT transition and the isotropic point coincide.

\section{Conclusions}

Our results reveal a novel promising probe for quantum phase transitions. By leveraging the textures formalism, an elegant window into the systems' dynamics, and then characterizing the system's texture by looking at the state rugosity, we were able to identify signatures at both phase transitions in the Heisenberg XXZ system, including the BKT transition at $\Delta = +1$. Remarkably, by using rugosity and sweeping across $\Delta$, we observe sharp and distinct features at the transition points, even for small sizes, from 6 spins onward. Our analytical connection between texture and the hierarchy of spin correlators shows that these features reflect the sensitivity of rugosity to changes in the correlation structure of the ground state. At $\Delta=1$, where the BKT transition coincides with the isotropic point of the underlying XXZ model, this sensitivity may provide a possible explanation for the absence of a discernible finite-size shift in the systems considered here. Whether this behavior is primarily associated with the enhanced symmetry at the isotropic point or more generally with BKT criticality remains an open question. We have reinforced this trend by looking at the finite-size scaling of the rugosity, which shows a clear dominance of the extensive contribution over the subleading corrections for the system sizes considered. We have also compared rugosity to existing probes, particularly the Bell measurement framework of Ref.~\cite{Justino2012}, which captures both transitions, but requires more experimental resources.

\begin{acknowledgments}

All authors thank Fernando Parísio for valuable discussions and feedback. H.P.C. thanks Nicolò Lo Piparo, Dario Poletti, Oliver Bellwood, Ivan Iakupov and Aoi Hayashi for discussions and suggestions. Computations were performed using the Deigo cluster at the Okinawa Institute of Science and Technology, and we acknowledge the Scientific Computing and Data Analysis Section at OIST. For the DMRG simulations, the ITensor library was used~\cite{fishman2022itensor}. The generated codes and data are available upon request. I.M.C. acknowledges support from the São Paulo Research Foundation (FAPESP), Grants No. 2023/00510-0 and No. 2025/05607-8. K.Z. acknowledges CNPq (Grant No.305665/2025-1). This work is supported in part by the MEXT Quantum Leap Flagship Program (MEXT Q-LEAP) under Grant No. JPMXS0118069605 and the Japan’s Council for Science, Technology and Innovation (CSTI) under the Cross-ministerial Strategic Innovation Promotion Program (SIP) for “Promoting the application of advanced quantum technology platforms to social issues” (JPJ012367).

\end{acknowledgments}

\section*{Texture and geometric measures of entanglement}

One can relate texture and rugosity with entanglement quantifiers by considering definitions based on distances between states. A particular example is a measurement known as global entanglement \cite{PhysRevA.71.060305} that is defined as the overlap between the state of interest and the closest separable state
\begin{equation}
    \Lambda_{\max}(\ket{\psi}) = \max_{\ket{\phi}} |\langle{\phi|\psi}\rangle|^2,
\end{equation}
where the maximization runs over all separable states $\ket{\phi} = \prod_j \ket{\phi_j}$.

One defines the measure of global entanglement as $\lambda(\ket{\psi}) = - \ln{\Lambda_{\max}}(\ket{\psi})$.

In order to draw a connection with the definition of texture, let's consider the computational basis, for which the flat state is itself a product state $\ket{f_1} = \prod_j \ket{+_j}$. since $|f_1\rangle$ is a specific product state, its overlap with $|\psi\rangle$ is bounded by the maximum overlap over all product states: $T(|\psi\rangle)$. Texture thus bounds the global entanglement by below
\begin{equation}
    \mathcal{T}(\ket{\psi}) \leq \Lambda_{\max}(\ket{\psi}),
\end{equation}
and,  conversely, rugosity is the upper bound of the geometric entanglement
\begin{equation}
    \lambda(\ket{\psi}) \leq \mathcal{R}(\ket{\psi}).
\end{equation}
This connection supports the interpretation of rugosity as an entanglement-related quantity.

\bibliography{refs}
\end{document}